\documentclass[journal]{/texlive/2016/texmf-dist/tex/latex/IEEEtran/IEEEtran}

\usepackage{cite}

\ifCLASSINFOpdf
   \usepackage[pdftex]{graphicx}
\else
\fi
\usepackage{amsmath}

\usepackage{amsfonts,amssymb}
\usepackage{subfigure}
\usepackage{booktabs}
\usepackage{multirow}
\usepackage{array}
\usepackage{pdfpages}
\begin{document}
%
\title{Jointly Adversarial Network to Wavelength Compensation and Dehazing of Underwater Images}
%
%

\author{Xueyan~Ding,
        Yafei~Wang,
        Yang~Yan,
        Zheng~Liang,
        Zetian~Mi,
        and~Xianping~Fu
\thanks{This work was supported in part by the National Natural Science Foundation of China Grant 61802043, Grant 61370142 and Grant 61272368, by the Fundamental Research Funds for the Central Universities Grant 3132016352, by the Dalian Science and Technology Innovation Fund 2018J12GX037 and Dalian Leading talent Grant, by the Foundation of Liaoning Key Research and Development Program. (Corresponding author: Xianping Fu)}
\thanks{The authors are with the school of Information Science and Technology, Dalian Maritime University, Dalian 116026, China (e-mail: dingxueyan@dlmu.edu.cn; wangyafei@mail.dlut.edu.cn; yanyang@dlmu.edu.cn; zliang@dlmu.edu.cn; mizetian@dlmu.edu.cn; fxp@dlmu.edu.cn).}
}

%
%

\markboth{Journal of \LaTeX\ Class Files,~Vol.~14, No.~8, August~2015}%
{Shell \MakeLowercase{\textit{et al.}}: Bare Demo of IEEEtran.cls for IEEE Journals}
%



\maketitle

\begin{abstract}
Severe color casts, low contrast and blurriness of underwater images caused by light absorption and scattering result in a difficult task for exploring underwater environments. Different from most of previous underwater image enhancement methods that compute light attenuation along object-camera path through hazy image formation model, we propose a novel jointly wavelength compensation and dehazing network (JWCDN) that takes into account the wavelength attenuation along surface-object path and the scattering along object-camera path simultaneously. By embedding a simplified underwater formation model into generative adversarial network, we can jointly estimates the transmission map, wavelength attenuation and background light via different network modules, and uses the simplified underwater image formation model to recover degraded underwater images. Especially, a multi-scale densely connected encoder-decoder network is proposed to leverage features from multiple layers for estimating the transmission map. To further improve the recovered image, we use an edge preserving network module to enhance the detail of the recovered image. Moreover, to train the proposed network, we propose a novel underwater image synthesis method that generates underwater images with inherent optical properties of different water types. The synthesis method can simulate the color, contrast and blurriness appearance of real-world underwater environments simultaneously. Extensive experiments on synthetic and real-world underwater images demonstrate that the proposed method yields comparable or better results on both subjective and objective assessments, compared with several state-of-the-art methods.

\end{abstract}


%
\IEEEpeerreviewmaketitle

\section{Introduction}
%
%
%
%

\begin{figure}[!t]
\centering
\includegraphics[width=3.4in]{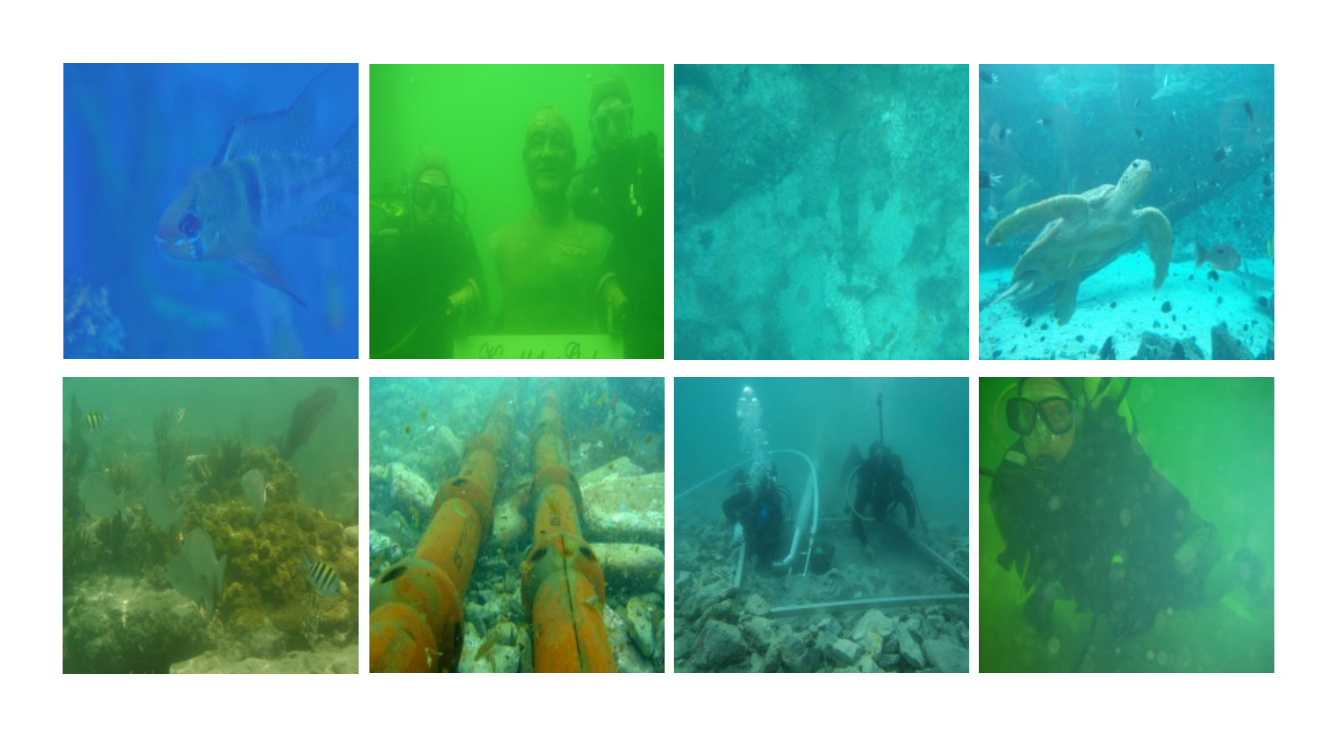}
\caption{Examples of underwater images with different color tones and degradation. These real-world underwater images are provided by previous methods and downloaded from the Internet.}
\label{fig_under-envir}
\end{figure}

\IEEEPARstart{E}{xploration} of the underwater environments such as biology and archaeology\cite{Ludvigsen2007Applications}, rescue operations or infrastructure inspection and maintenance\cite{Bonin2011Imaging}, can highly benefit from underwater vehicles equipped with optical imaging sensors. Unfortunately, as shown in Fig. \ref{fig_under-envir}, underwater optical imaging shows some degradation effects such as color casts, blurriness and low contrast. As light travels in the water, it is exponentially attenuated through two processes: absorption and scattering. The two effects are due not only to the water itself but also to other components such as dissolved organic matter or small observable floating particles \cite{Schettini2010Underwater}. Absorption effect reduces the amount of ambient light from the longest wavelength to shortest. The red light first disappears, followed by green light and blue light, thereby making the underwater images to be dominated essentially by green and blue color. Absorption effect causes important consequences: altering the color perception of the scene and limiting the distance at which objects are perceived. Scattering appears as a deviation of a photon from its original ways after hitting a particle suspended in the water. Depending on the deviation angle of the light ray, this phenomenon is known as forward scattering or backward scattering. Forward scattering randomly occurs when the light travels from the object to the camera, resulting in a blur scene. On the other hand, backward scattering occurs when the light is reflected to the camera before it reaches the objects, leading to a limit of contrast and generating a veil that superimposes itself on the scene. In addition, light attenuation (absorption and scattering) limits the visibility range at around twenty meters in clear water and five meters or less in turbid water\cite{Schettini2010Underwater}. The visibility range can be increased with artificial lighting, but these sources leads to a non-uniform illuminated scene that presents a bright spot in the image with a poorly illuminated area surrounding it. Affected by these properties, underwater optical imaging is limited when being used for display and extracting valuable information for further processing. Therefore, an effective method that can tackle these problems for both display and analysis is meaningful, and thus desired.

In this paper, we propose a jointly wavelength compensation and dehazing strategy via embedding a simplified underwater image formation model into generative adversarial network framework, to remove color casts and scattering simultaneously.
This paper makes the following contributions:

\begin{itemize}
\item
A novel jointly optimizable wavelength compensation and dehazing network to reconstruct the clear underwater image while preserving the original structure and texture is proposed. This is achieved by directly embedding an underwater image formation model that considers the wavelength attenuation along the surface-object path and scattering along the object-camera path, into generative adversarial network framework via math operation modules. It allows a jointly learning task for estimating the transmission map, wavelength attenuation and background light, and reconstructing the clear underwater image.
\item
A novel inherent optical properties (IOPs, the absorption and scattering coefficients in particular) based underwater image synthesis method that can simulate the wavelength attenuation and scattering simultaneously for different underwater types is proposed. The inherent optical properties represent the color, contrast and blurriness appearance of real-world underwater environments. To our best knowledge, it is the first underwater image synthesis method that takes into account the wavelength attenuation and scattering simultaneously for different underwater types.

\item
A multi-scale densely connected encoder-decoder network with a gradient preserving loss function is proposed for accurately estimating the transmission map, in order to maintain sharp edges and avoid halo artifacts when dehazing. The proposed network consists of densely connected module and multi-scale module, which can make use of the features from multiple layers of a CNN.

\item
Extensive experiments are conducted on one synthetic dataset and two real-world image datasets. In addition, comparisons are performed against several recent state-of-the-art methods.

\end{itemize}


\section{RELATED WORK}
Various methods for degraded underwater images have been developed in the past years. These methods are usually depend on single image or multiple images. Methods such as \cite{Schechner2004Clear} and \cite{Sheinin2015The}, take advantage of the additional information of multiple images. These images are usually captured by polarization filters or specialized hardware devices. Capturing multiple images with specialized devices and using these images can be expensive and time-consuming, which reduce their applicability. Besides, the underwater image processing relying on single image is usually addressed from two different viewpoints: as an image restoration method or as an image enhancement method.

\begin{figure*}[!t]
\centering
\includegraphics[width=7.1in]{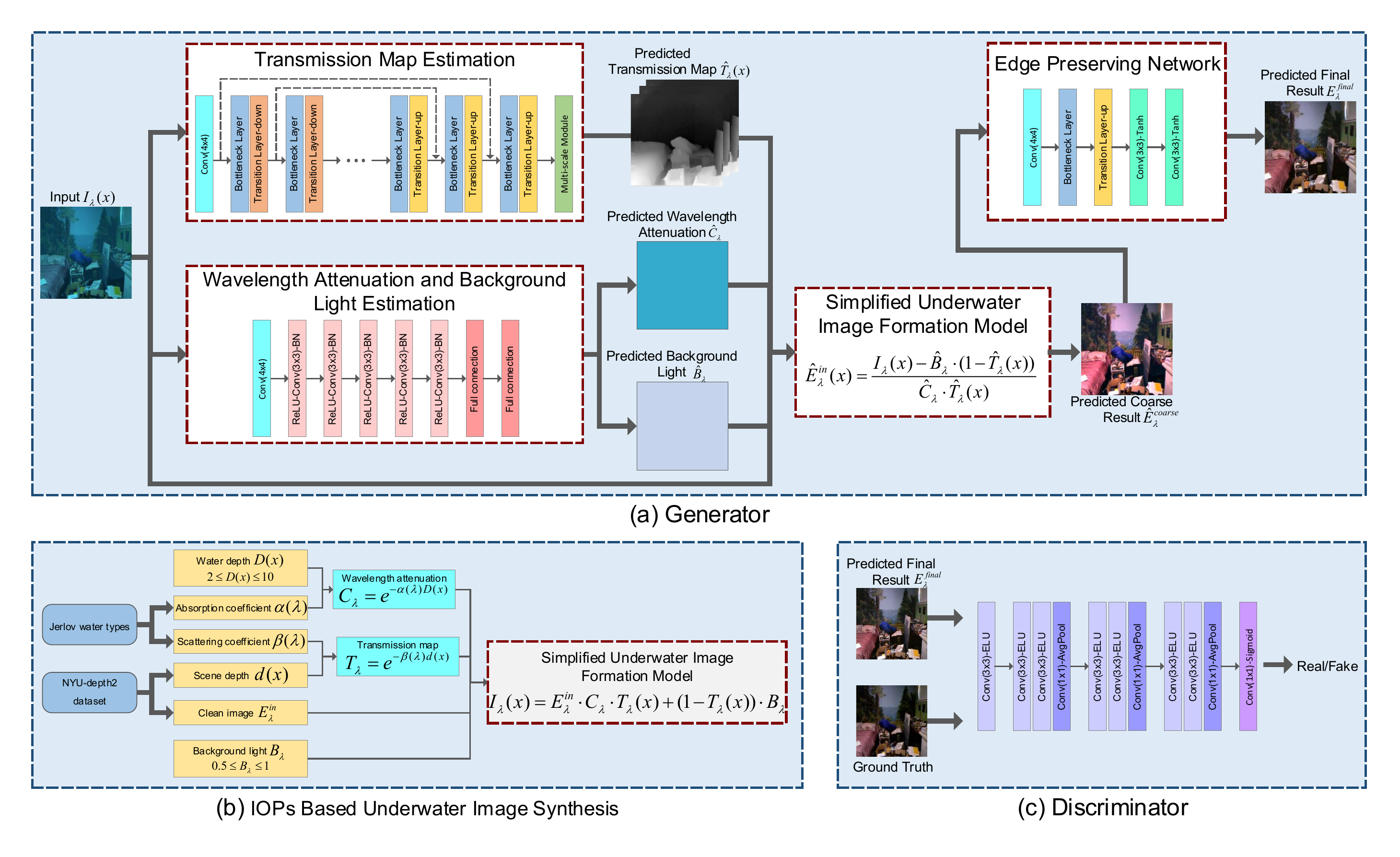}
\caption{An overview of the proposed method. (a) and (c) show the generator and discriminator of the proposed JWCDN underwater image restoration method. (b) shows the proposed IOPs based underwater image synthesis method.}
\label{fig_overall}
\end{figure*}

Image restoration, constructing a physical model of the degradation of images to recover degraded images, is essentially an inverse problem. The image restoration methods need to estimate many model parameters such as attenuation and diffusion coefficients that are extremely variable and scarcely known. Jaffe-McGlamery model \cite{McGlamery1979Acomputer} \cite{Jaffe1990Computer} is a distinguished underwater image formation model that decomposes an underwater image into three major components: direct, forward-scatter, and backscatter components. Based on this model, \cite{Zhao2015Deriving} extracted inherent optical properties of water from the background color of degraded underwater images to remove the degradation effects. Considering the similar hazy appearance between underwater scenes and hazy scenes, simplified hazy image formation model is also applied widely in poor underwater conditions, called underwater image dehazing. \cite{Chiang2012Underwater} combined the simplified hazy image formation model with a wavelength dependent compensation algorithm that aimed to remove the bluish tone of degraded underwater images. However, this method is limited to the underwater images with serious color casts. Built on a minimum information loss principle, \cite{Li2016Underwater} proposed an underwater image dehazing algorithm combined with a histogram distribution prior to restore the visibility, color and contrast. Additionally, \cite{Drews2016Underwater} modified the dark channel prior presented in \cite{He2011Single}, called UDCP, to improve degraded underwater images. However, UDCP does not always hold when there are white objects or artificial light in the underwater scenes. Similarly,\cite{Peng2018Generalization} introduced a generalized dark channel prior that calculates the difference between the observed intensity and the ambient light to obtain the transmission map. After that, the degraded images are recovered according to an image formation model. Moreover, \cite{Galdran2015Automatic} proposed a red channel based contrast recovery method via restoring the colors associated with short wavelengths. \cite{Peng2017Underwater} investigated a single image oriented algorithm to restore underwater images based on the image blurriness and light absorption (IBLA). Additionally, \cite{Berman2018Underwater} transformed underwater image enhancement into image dehazing with the attenuation ratios of the blue-red and blue-green color channels.

Image enhancement methods adjust the pixel values of a image without relying on any physical imaging model to generate a more visually pleasing image. The enhancement results are qualitative and subjective. Traditional filters, histogram equalization based methods and white balance methods are unable to adaptively improve the color casts, low contrast and blurriness of underwater images, due to the changeability of underwater environments. \cite{Fu2014Retinex} proposed a novel retinex-based enhancing approach to enhance single underwater image. They use a variational framework for retinex to decompose the reflectance and the illumination, and enhance the reflectance and the illumination to address the under-exposure and fuzz problem. \cite{Ancuti2012Enhancing} proposed a fusion-based structure to improve the visibility of underwater images and videos. They used weights to fuse a contrast improved version and a color corrected version derived from a degraded underwater image. Recently, \cite{Ancuti2018Color} modified this previous work, in order to improve the color appearance of resultant images and reduce the effects of the over-exposure. \cite{Ghani2015Underwater}, \cite{Ghani2015Enhancement} presented a Rayleigh-stretched contrast-limited adaptive histogram method. This method reduces the amount of under-enhanced and over-enhanced regions, but tends to increase noise in the resultant images. \cite{Li2017A} proposed a hybrid method based on image color prior and dehazing for underwater image enhancement. The method corrects the color casts of underwater image and improves the visibility. This method shows limitations when the image color prior is not available. Convolutional networks have been recently used for dehazing \cite{Ren2016Single, Cai2016Dehazenet, Li2017An, Zhang2018Densely}, and also for underwater restoration \cite{Li2017WaterGAN,Fabbri2018Enhancing}. \cite{Li2017WaterGAN} proposed a CNN-based real-time underwater image color correction network, called WaterGAN. The model is trained by synthetic underwater images. However, this method is only effective for underwater images captured under specific circumstances which are similar to its training data. \cite{Fabbri2018Enhancing} proposed an underwater image technique based on Generative Adversarial Networks for autonomous underwater robots. The model is trained by synthetic underwater images as well. \cite{Li2018Emerging} presented an underwater image color correction method using weakly supervised color transfer. It relaxes the demand of paired underwater images for training by learning a mapping function between underwater images and air images. But it highly relies on training images. Although deep learning has achieved impressive performance on low-level vision tasks, there is still a limit of the development of deep learning based underwater image enhancement methods due to the insufficiency of labeled training data.

\section{PROPOSED METHOD}
To remove color casts and scattering simultaneously, we introduce a jointly compensating wavelength and dehazing strategy via embedding a simplified underwater image formation model into generative adversarial network framework. Fig. \ref{fig_overall}(a) and Fig. \ref{fig_overall}(c) show the generator and discriminator of the proposed JWCDN underwater image restoration method. For the above jointly learning tasks, we propose an IOPs based underwater image synthesis method as shown in Fig. \ref{fig_overall}(b). In what follows, we explain these modules in detail.

\subsection{Simplified Underwater Image Formation Model}

\begin{figure}[!t]
\centering
\includegraphics[width=3.4in]{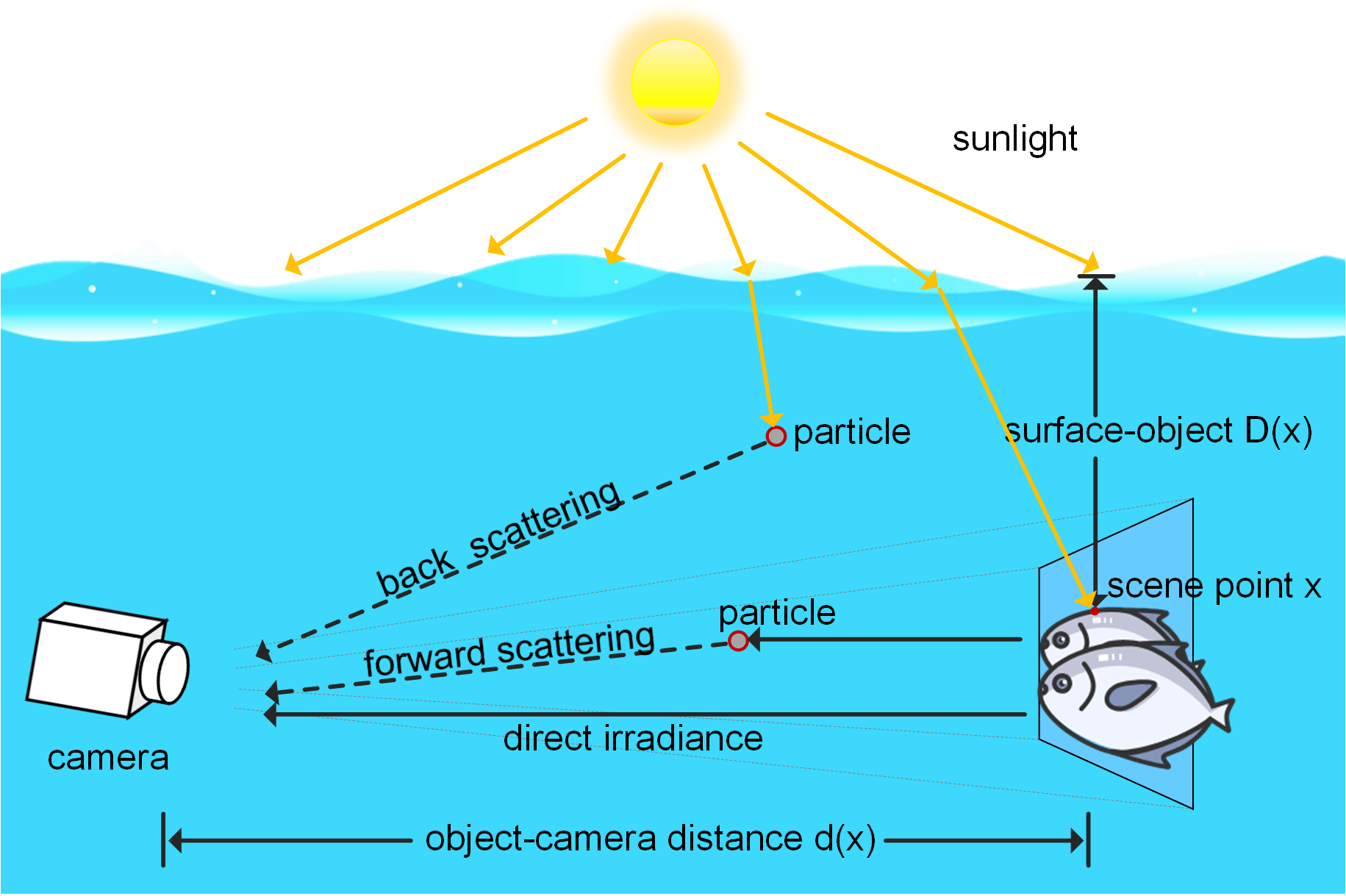}
\caption{Schematic diagram of underwater optical imaging. Natural light enters from air to an underwater scene point $x$ with water depth $D(x)$. Then the light reflected propagates distance $d(x)$ to the camera. The irradiance perceived by camera is suffered from two factors: light scattering and light absorption.}
\label{fig_undermodel}
\end{figure}

In Fig.\ref{fig_undermodel}, homogeneous skylight firstly travels from the surface of water and reaches the underwater scene point $x$ with water depth $D(x)$, and then the light reflected propagates distance $d(x)$ to the camera. According to the Lambert-Beer empirical law, the degradation of light intensity is relevant to the properties of the material which light travels through by an exponential dependence. The energy attenuation when light propagates the surface-object path $D(x)$ and reaches the underwater scene point $x$, can be expressed as
\begin{eqnarray}
C_{\lambda} = e^{-\alpha(\lambda)D(x)}
\label{Eqs-Clambda}
\end{eqnarray}
where $\alpha(\lambda)$ indicates the spectral absorption coefficient of the medium that depends on the wavelength $\lambda$. $C_{\lambda}$ represents the wavelength attenuation coefficient which is a primary cause of color casts. The irradiance of the object $E_{\lambda}^{obj}$ in the water can be modeled as
\begin{eqnarray}
E_{\lambda}^{obj} = E_{\lambda}^{in}\cdot C_{\lambda}
\label{Eqs-obj}
\end{eqnarray}
where $E_{\lambda}^{obj}$ represents the irradiance of the object due to vertical absorption in light of the initial energy $E_{\lambda}^{in}$. During the course of surface-object propagation, light with different wavelengths is subjected to varying degrees of attenuation, causing color change of underwater images. It is noteworthy that wavelength attenuation is actually along both the surface-object path and object-camera path. But in this paper, we don't take into account the wavelength attenuation occurred in object-camera path for computational efficiency.

To describe the propagation of object-camera path, Researchers\cite{Drews2016Underwater}\cite{Peng2018Generalization} usually use a simplified hazy image formation model
\begin{eqnarray}
\begin{aligned}
I_{\lambda}(x) = J_{\lambda}(x)\cdot T_{\lambda}(x) + (1-T_{\lambda}(x))\cdot B_{\lambda}, \\
\lambda \in \{red,green,blue\}
\label{Eqs-hazy}
\end{aligned}
\end{eqnarray}
where $x$ is a point in the underwater scene, $I_{\lambda}(x)$ is the total irradiance detected by the camera, i.e. the light reflected again travels distance $d(x)$ to the camera forming pixel $I_{\lambda}(x)$, $\lambda \in \{red,green,blue\}$.
$J_{\lambda}(x)$ is the scene irradiance at point $x$, $T_{\lambda}(x)$ is the transmission map, $B_{\lambda}$ is the homogeneous background light, and $\lambda$ is the light wavelength. This equation is achieved by approximating the true in-scattering term in the full radiative transport equation with the background light in an underwater image.

\begin{figure}[!t]
\centering
\includegraphics[width=3.45in]{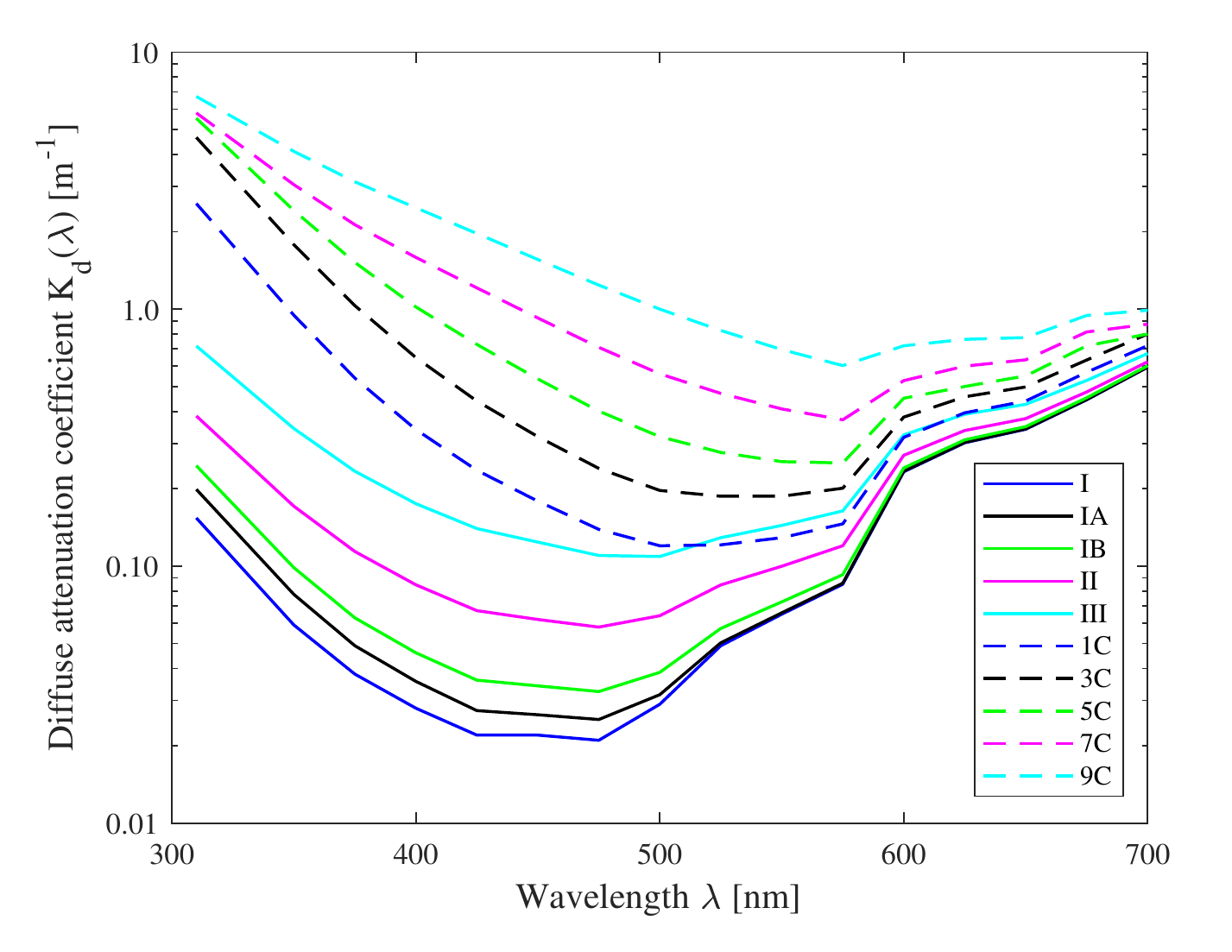}
\caption{Diffuse attenuation coefficients $K_{d}(\lambda)$ of Jerlov water types from \cite{Jerlov1968Irradiance}. Solid lines mark open ocean water types (I through III) while dashed lines mark coastal water types (1C through 9C). For open ocean waters, Type-I is the clearest and Type-III is the most turbid. Similarly, Type-1C is clearest coastal water and Type-9C is the most turbid.}
\label{fig_Jerlov}
\end{figure}

\renewcommand\arraystretch{1.2}
\begin{table*}[]
\scriptsize
\centering
\caption{Absorption $\alpha(\lambda)$ and Scattering $\beta(\lambda)$ values of Jerlov water types \cite{Solonenko2015Inherent}.}
\label{Tab_Jerlov}
\begin{tabular}{p{0.6cm} p{0.4cm}p{0.6cm} p{0.4cm}p{0.6cm} p{0.4cm}p{0.5cm} p{0.4cm}p{0.5cm} p{0.4cm}p{0.35cm} p{0.4cm}p{0.5cm} p{0.4cm}p{0.4cm} p{0.4cm}p{0.4cm} p{0.4cm}p{0.35cm} p{0.4cm}p{0.35cm}<{\centering}}
\toprule
\multirow{3}{*}{\shortstack{Wave-\\length,\\nm}}
& \multicolumn{2}{c}{I}  & \multicolumn{2}{c}{IA} & \multicolumn{2}{c}{IB} & \multicolumn{2}{c}{II} & \multicolumn{2}{c}{III}
& \multicolumn{2}{c}{1C} & \multicolumn{2}{c}{3C} & \multicolumn{2}{c}{5C} & \multicolumn{2}{c}{7C} & \multicolumn{2}{c}{9C}\\

\cmidrule(r){2-3}   \cmidrule(r){4-5}   \cmidrule(r){6-7}   \cmidrule(r){8-9}   \cmidrule(r){10-11}
\cmidrule(r){12-13} \cmidrule(r){14-15} \cmidrule(r){16-17} \cmidrule(r){18-19} \cmidrule(r){20-21}

& $\alpha(\lambda)$ & $\beta(\lambda)$ & $\alpha(\lambda)$ & $\beta(\lambda)$ & $\alpha(\lambda)$ & $\beta(\lambda)$ & $\alpha(\lambda)$ & $\beta(\lambda)$ & $\alpha(\lambda)$ & $\beta(\lambda)$
& $\alpha(\lambda)$ & $\beta(\lambda)$ & $\alpha(\lambda)$ & $\beta(\lambda)$ & $\alpha(\lambda)$ & $\beta(\lambda)$ & $\alpha(\lambda)$ & $\beta(\lambda)$ & $\alpha(\lambda)$ & $\beta(\lambda)$ \\

\midrule

450	&	0.018	&	0.0038	&	0.022	&	0.0063	&	0.024	&	0.062	&	0.024	&	0.504	&	0.039	&	1.38	&	0.105	&	0.514	&	0.154	&	1.5	&	0.297	&	1.87	&	0.542	&	3.3	&	0.943	&	4.39	\\

525	&	0.046	&	0.0021	&	0.047	&	0.0040	&	0.047	&	0.078	&	0.047	&	0.387	&	0.051	&	1.06	&	0.068	&	0.395	&	0.078	&	1.15	&	0.127	&	1.44	&	0.233	&	2.54	&	0.43	&	3.38	\\

650	&	0.334	&	0.0009	&	0.334	&	0.0023	&	0.334	&	0.393	&	0.334	&	0.27	&	0.336	&	0.74	&	0.344	&	0.274	&	0.346	&	0.8	&	1.78	&	2.87	&	0.403	&	1.77	&	0.456	&	2.35	\\
\bottomrule
\end{tabular}
\footnotesize{The units of $\alpha(\lambda)$ and $\beta(\lambda)$ are $\textup m^{-1}$. }\\
\end{table*}

\begin{figure*}[!t]
\centering
\includegraphics[width=7.2in]{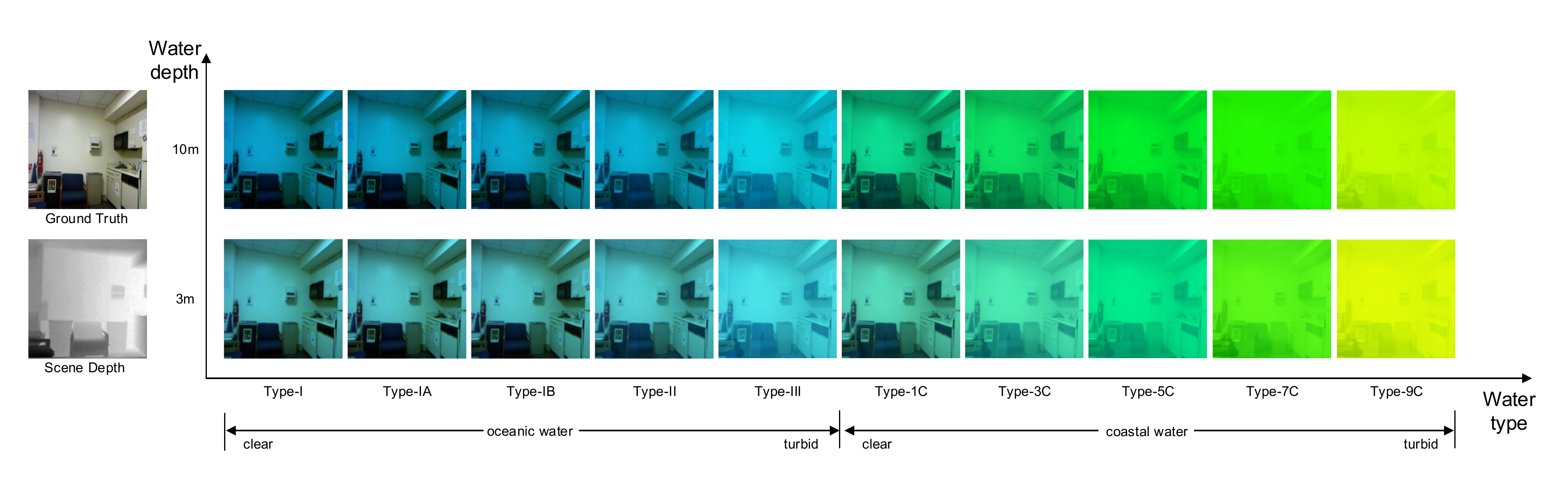}
\caption{Samples of synthesized underwater images with different types and water depths generated from NYU-depth2 dataset\cite{Silberman2012Indoor}. We show synthesized underwater images where water depth are 3m and 10m, and choose a random homogeneous global background light $0.5\leq B_{\lambda}\leq 1$.}
\label{fig_generate}
\end{figure*}

Additionally, there is an exponential relationship between the transmission $T_{\lambda}(x)$ and the scene depth $d(x)$, i.e.
\begin{eqnarray}
T_{\lambda}(x) = e^{-\beta(\lambda)d(x)}
\label{Eqs-Tlambda}
\end{eqnarray}
where $\beta(\lambda)$ is the scattering coefficient. And light $J_{\lambda}(x)$ emanated from point $x$ is equal to the amount of illuminating ambient light $E_{\lambda}^{obj}$ reflected, i.e.
\begin{eqnarray}
\begin{aligned}
J_{\lambda}(x)&= E_{\lambda}^{obj}\cdot \rho_{\lambda}(x) \\
&=E_{\lambda}^{in}\cdot C_{\lambda} \cdot \rho_{\lambda}(x)
\label{Eqs-rho}
\end{aligned}
\end{eqnarray}
where $\rho_{\lambda}(x)$ represents the reflectivity coefficient of point $x$ for light with wavelength $\lambda$. Assuming $\rho_{\lambda}(x)=1$, a simplified underwater image formation model which is a combination of the courses of surface-object and object-camera propagation, can be modeled as

\begin{eqnarray}
\begin{aligned}
I_{\lambda}(x) = (E_{\lambda}^{in}\cdot C_{\lambda})\cdot T_{\lambda}(x) + (1-T_{\lambda}(x)) \cdot B_{\lambda},\\
\lambda \in \{red,green,blue\}
\label{Eqs-under1}
\end{aligned}
\end{eqnarray}

Once the transmission map, the background light and the wavelength attenuation are estimated, underwater images can be recovered as follows
\begin{eqnarray}
\begin{aligned}
\hat E_{\lambda}^{in} = \frac{I_{\lambda}(x)-\hat B_{\lambda}(1-\hat T_{\lambda}(x))}{\hat C_{\lambda}\cdot \hat T_{\lambda}(x)},\lambda \in \{red,green,blue\}
\label{Eqs-under2}
\end{aligned}
\end{eqnarray}

\subsection{IOPs Based Underwater Image Synthesis}

Unlike the high-level visual analytics tasks where large training datasets are often available, the learning-based underwater image enhancement task is lack of training datasets. The underwater image synthesis is usually addressed from two different viewpoints: based on physical model\cite{Anwar2018Deep} or CycleGAN\cite{Fabbri2018Enhancing} \cite{Zhu2017Unpaired}. Physical model based synthesis usually relies on hazy formation model, which concentrates on the light attenuation along object-camera path and neglects the wavelength attenuation that increases with water depth. There are some color shifts between the synthesized underwater images generated by hazy formation model and real-world underwater images. CycleGAN based synthesis shows a great advantage for unpaired data, enabling to transfer style of images instead of all color casts, low contrast and blurriness of images. But CycleGAN based synthesis highly depends on training samples, and can not satisfy the demand of multiple parameters of training samples.

We introduce an IOPs based underwater image synthesis method using the simplified underwater image formation model which describes the wavelength attenuation along surface-object path and light scattering along object-camera path. Considering the demand of multiple parameters of training samples for the proposed network, we establish a dataset \{\textit{Degraded/Clean/Transmission Map/Wavelength Attenuation/Background Light}\}. Different from other physical model based synthesis methods that generate training set with a random sample of scattering coefficients, we introduce the inherent optical properties (absorption and scattering coefficients) of different water types into synthesis process. Table \ref{Tab_Jerlov} shows the absorption $\alpha(\lambda)$ and scattering $\beta(\lambda)$ that presented in \cite{Solonenko2015Inherent} for all Jerlov water types \cite{Jerlov1968Irradiance} over the wavelength 650nm (red), 525nm (green) and 450nm (blue). Jerlov categorized waters into five distinct oceanic types (I through III) and five coastal types (1C through 9C) as shown in Fig. \ref{fig_Jerlov}. For oceanic waters, Type-I is the clearest and Type-III is the most turbid. Likewise, Type-1C is clearest coastal water and Type-9C is the most turbid.

\begin{figure*}[!t]
\centering
\includegraphics[width=7.1in]{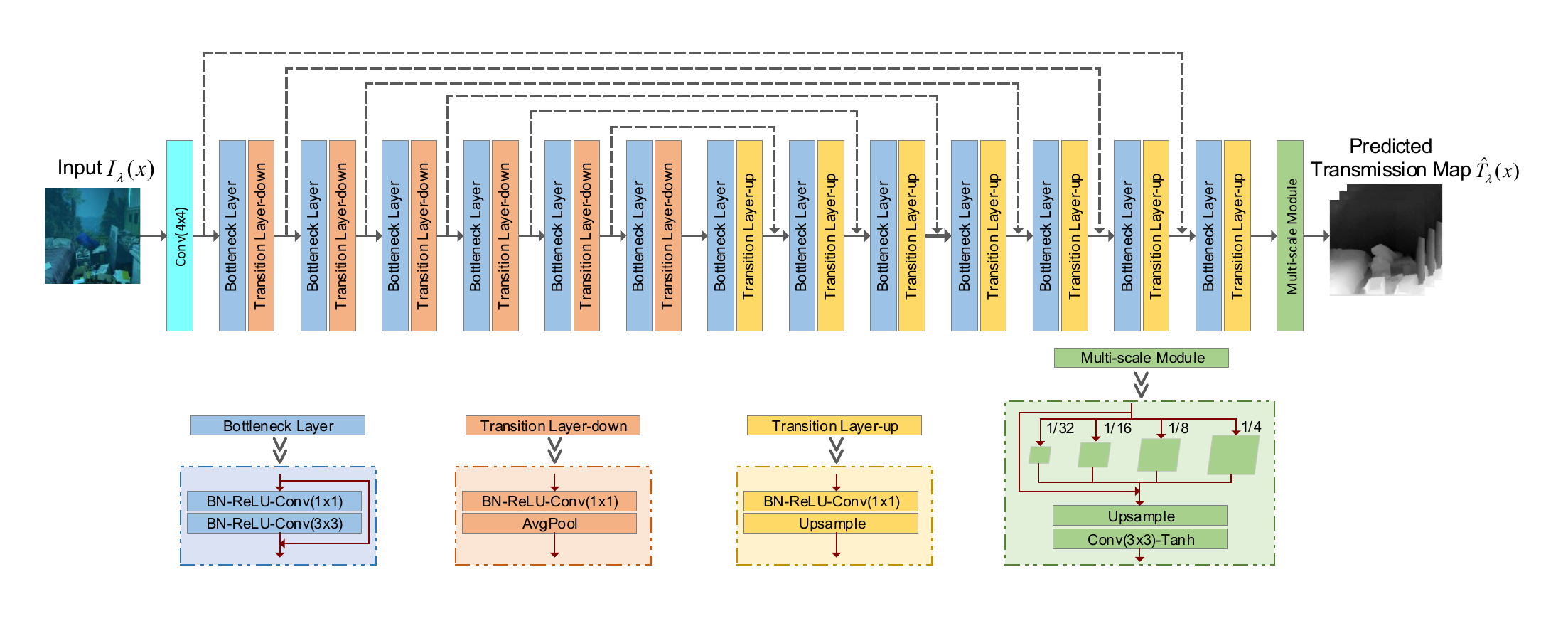}
\caption{An overview of the proposed multi-scale densely connected transmission map estimation network.}
\label{fig_network}
\end{figure*}

We establish the underwater image dataset for different water types using the NYU-depth2 dataset\cite{Silberman2012Indoor}, which contains clean indoor images and their corresponding scene depths. Specifically, the clean indoor images in NYU-depth2 are regarded as the parameter \{\textit{Clean}\} of the multiple parameters set. And randomly selecting a water type, the corresponding $\alpha(\lambda)$ and $\beta(\lambda)$ values of the red (650nm), green (525nm), and blue (450nm) channels for the water type can be derived from Table \ref{Tab_Jerlov}. Based on absorption coefficient $\alpha(\lambda)$, the color-cast $C_{\lambda} = e^{-\alpha(\lambda)D(x)}$ can be generate with the water depth $2\leq D(x)\leq 10$. Transmission map $T_{\lambda}(x) = e^{-\beta(\lambda)d(x)}$ can be generate with the scattering coefficient $\beta(\lambda)$ and scene depth from NYU-depth2 dataset. Moreover, we select a random homogeneous global background light $0.5\leq B_{\lambda}\leq 1$, and apply Eq. \ref{Eqs-under1} to generate degraded underwater images. The process of the proposed IOPs based underwater image synthesis method is shown in Fig. \ref{fig_overall}(b). In addition, Fig. \ref{fig_generate} shows sample synthesized underwater images with different types and water depths.

We can train the proposed jointly wavelength compensation and dehazing network with the synthesized underwater image dataset. In what follows, we explain the network in detail.

\subsection{Network Architecture}

The proposed network architecture is composed of four modules: 1) multi-scale densely connected transmission map estimation net, 2) wavelength attenuation and background light estimation net, 3) wavelength compensation and dehazing via underwater image formation model, and 4) edge preserving net. A flowchart of the proposed network architecture is illustrated in Fig. \ref{fig_overall}(a) (c).

\subsubsection{Multi-scale Densely Connected Transmission Map Estimation Network}
Inspired by previous methods that use multi-scale features to estimate the transmission map\cite{Ren2016Single, Cai2016Dehazenet,Li2017An}, we propose a novel densely connected network to predict transmission map $T_{\lambda}$, where a degraded underwater image is the input. The network consists of two modules, i.e. densely connected module and multi-scale module. Fig. \ref{fig_network} shows an overview of the proposed multi-scale densely connected transmission map estimation network.

The densely connected module contains layers of symmetric encoder and decoder where the dense block \cite{Huang2017Densely} is used as the basic structure. Specifically, we adopt a convolutional layer and six Bottleneck Layers with their corresponding down-sampling Transition Layers as the encoder structure. The Bottleneck Layer corresponds the sequence BN-ReLU-Conv($1\times 1$)-BN-ReLU-Conv($3\times 3$), and the down-sampling Transition Layer corresponds the sequence BN-ReLU-Conv($1\times 1$)-AvgPool \footnote{Conv:Convolution, BN: Batch-normalization\cite{Ioffe2015Batch} and AvgPool: Average pool.}. At end of the encoding structure, the feature size is 1/128 of the input size. And we stack seven Bottleneck Layers with the refined up-sampling Transition Layers \cite{Jégou2017Fully,Zhu2017Densenet} as the decoder structure, in order to reconstruct the features into original resolution. Additionally, we apply dense connections to build short-cut paths between upstream and downstream convolutional blocks, which enable the model to connect the convolutional feature maps to the deconvolutional feature maps with the same dimension, thus produce a larger scale of features. Dense blocks and skip shortcuts improve the information flow along those features and guarantee better convergence via connecting all layers.

Apart from the combination of different features achieved by the densely connected structure, we make use of the "global" structural features of objects with different scales extracted by a multi-scale pyramid pooling module \cite{Zhang2018Densely} \cite{Zhao2017Pyramid}. The reason is that the result generated by just densely connected encoder-decoder structure is lack of the structural features of objects with different scales. Rather than taking very large pooling size to capture more global context information between different objects, more "local" information to characterize the "global" structure of each object is needed \cite{Zhang2018Densely} \cite{Zhao2017Pyramid}. Specifically, we adopt a four-scale pooling operation where pooling sizes are 1/32, 1/16, 1/8 and 1/4 of the original size. Then, we up-sample all four scale features to original feature size and concatenate them with the original feature before the final estimation.

\subsubsection{Wavelength Attenuation and Background Light Estimation Network}
Following the simplified underwater image formation model Eq. \ref{Eqs-under1}, wavelength attenuation $C_{\lambda}$ is an RGB vector as well as background light $B_{\lambda}$. We concatenate wavelength attenuation and background light as a vector, and estimated it as a classification task. We adopt a convolutional layer, five ReLU-Conv($3\times 3$)-BN blocks that similar to basic block of U-Net \cite{Ronneberger2015Unet}, and two fully-connected layers as our network structure.


\subsubsection{Underwater Image Recovery via the simplified underwater image formation model}
To bridge the relation among the transmission map, the background light, wavelength attenuation and the recovered image, and to make sure that the whole network structure is jointly optimized for all three tasks, we directly embed the simplified underwater image formation model into the overall optimization framework. Based on the jointly optimized tasks and the embedded model, a coarse recovered result $E_{\lambda}^{coarse}$ can be obtained.
\subsubsection{Edge Preserving Network}
We adopt an edge preserving network to refine the coarse prediction $E_{\lambda}^{coarse}$ and obtain a better recovered result $E_{\lambda}^{final}$, to prevent the possible blurriness which exists in real-world underwater images or caused by the three tasks aforementioned. The edge preserving network consists of a convolutional layer, a Bottleneck Layer, a Transition Layer and two Conv($3\times3$)-Tanh blocks.

\subsection{Formulation}
Our goal is to learn  mapping functions between two domains $X$ and $Y$ given training samples $\{I_{\lambda}^{i}\}_{i=1}^{N}$ where $I_{\lambda}\in X$ and $\{J_{\lambda}^{j}\}_{j=1}^{M}$ where $J_{\lambda}\in Y$. We denote the data distribution as $I_{\lambda}\sim p_{data}(I_{\lambda})$ and $J_{\lambda}\sim p_{data}(J_{\lambda})$. Let $G_{t}$, $G_{cb}$ denote the networks that generate the transmission map, the wavelength attenuation and background light, respectively. And let $G$ denotes the generator of JWCDN. To make sure that the estimated transmission map $G_{t}(I_{\lambda})$, wavelength attenuation and background light $G_{cb}(I_{\lambda})$, and the final recovered image $G(I_{\lambda})$ are indistinguishable from their corresponding ground truths, respectively, we make use of the following losses.

\subsubsection{Adversarial Loss}
We applied adversarial loss \cite{Goodfellow2014Generative} to the mapping function. For the mapping function $G: X \rightarrow Y$ and its discriminator $D$, we express the objective as
\begin{eqnarray}
\begin{aligned}
\mathcal{L}_{adv}&=\mathbb{E}_{J_{\lambda}\sim p_{data}(J_{\lambda})}[\log D(J_{\lambda})]\\
&+\mathbb{E}_{I_{\lambda}\sim p_{data}(I_{\lambda})}[\log (1-D(G(I_{\lambda})))]
\label{Eqs-Loss}
\end{aligned}
\end{eqnarray}
where generator $G$ tries to generate samples $G(I_{\lambda})$ that look similar to real samples $J_{\lambda}$, while discriminator $D$ aims to distinguish between generated samples $G(I_{\lambda})$ and real samples $J_{\lambda}$. $G$ aims to minimize this objective against the adversary $D$ that tries to maximize it, i.e. $\min_{G}\max_{D}\mathcal{L}_{adv}$. The generator $G$ is trained to generate images that cannot be distinguished from "real" images by an adversely trained discriminator $D$, while the discriminator $D$ is trained to distinguish the generator's "fakes" as well as possible.

\subsubsection{Content Preserving Loss}
Pixel-wise loss such as L1 loss and L2 loss are generally used to help to recover the low frequency details of the generated image. Considering L1 loss encourages less blurring that L2 \cite{Isola2017Image}, we employ L1 loss between $G(I_{\lambda})$ and real samples $J_{\lambda}$. Apart from the low frequency part, we also use a perceptual loss \cite{Johnson2016Perceptual} to ensure the correctness of the high frequency part of the generated image $G(I_{\lambda})$. Thus, we leverage a weighted combination of a pixel-based loss and a feature-based loss. It is defined as following

\begin{eqnarray}
\begin{aligned}
\mathcal{L}_{cont} &= \omega_{p}\cdot \left\| G(I_{\lambda})-J_{\lambda}\right\|_1 \\
&+ \omega_{f}\cdot \left\| \phi(G(I_{\lambda}))-\phi(J_{\lambda})\right\|_1
\label{Eqs-content}
\end{aligned}
\end{eqnarray}
where $\left\| G(I_{\lambda})-J_{\lambda}\right\|_1$ is L1 loss and also a per-pixel loss, and $\left\|\phi(G(I_{\lambda}))-\phi(J_{\lambda})\right\|_1$ is a feature-based loss that calculates the difference of CNN feature maps between the generated image $G(I_{\lambda})$ and target image $J_{\lambda}$. $\phi$ denotes a feature extractor from the layers before 1st and 2th pooling layers in VGG16 network.

\subsubsection{Gradient Preserving Loss}
Inspired by the gradient loss used in depth estimation \cite{Ummenhofer2017Demon} and image enhancement \cite{Fabbri2018Enhancing}, we adopt a gradient loss to maintain the gradient of the generated transmission map $G_{t}(I_{\lambda})$. The gradient loss can be expressed as
\begin{eqnarray}
\begin{aligned}
\mathcal{L}_{grad} = \sum\limits_{w,h}\left\| (H_{x}(G_{t}(I_{\lambda})))_{w,h}-(H_{x}(T_{\lambda}))_{w,h}\right\|_2 \\
+ \sum\limits_{w,h} \left\| (H_{y}(G_{t}(I_{\lambda})))_{w,h}-(H_{y}(T_{\lambda}))_{w,h}\right\|_2
\label{Eqs-gradient}
\end{aligned}
\end{eqnarray}
where $H_{x}$ and $H_{y}$ are operators that compute image gradients along rows (horizontal) and columns (vertical) respectively, and $w\times h$ denotes the width and height of the transmission map. Additionally, we use a L1 loss to maintain the content of generated transmission map $G_{t}(I_{\lambda})$. Hence, the objective of transmission map estimation network $G_{t}$ can be expressed as
\begin{eqnarray}
\begin{aligned}
\mathcal{L}_{t} = \mathcal{L}_{grad} + \left\| G_{t}(I_{\lambda})-T_{\lambda}\right\|_1
\label{Eqs-trans}
\end{aligned}
\end{eqnarray}

\subsubsection{Full Objective}
The objective function of our method consists of four loss functions
\begin{eqnarray}
\mathcal{L} = \mathcal{L}_{adv} + \mathcal{L}_{cont} + \mathcal{L}_{t} + \mathcal{L}_{cb}
\label{Eqs-overall}
\end{eqnarray}
where $\mathcal{L}_{cb}$ is composed of $L1$ loss in predicting the wavelength attenuation and background light.

\section{EXPERIMENTAL RESULTS}
To evaluate our method, we perform qualitative and quantitative comparisons on both synthetic and real-world underwater images. All the results are compared with several state-of-the-art underwater image enhancement methods: Retinex-based\cite{Fu2014Retinex}, Histogram Prior\cite{Li2016Underwater}, GDCP\cite{Peng2018Generalization}, UDCP\cite{Drews2016Underwater}, IBLA\cite{Peng2017Underwater}, and UGAN\cite{Fabbri2018Enhancing}. We run the source codes provided by the corresponding authors with the recommended parameter settings to produce the best results for an objective evaluation. Since WaterGAN\cite{Li2017WaterGAN} is only applicable to specific circumstances, its results are not competitive and therefore not reported in this paper. In addition, we conduct an ablation study to demonstrate the effectiveness of each module of our network.

\subsection{Implementation Details}

We divide the NYU-depth2 dataset (1449 images) into three groups which contains 1000, 100 and 349 images separately, in order to generate a training set, a validation set and a test set. We choose four water types, i.e. Type-II, Type-III, Type-1C and Type-3C, to train the proposed network, considering the other water types are either too clear or too turbid. And the chosen four types can satisfy the general demand of underwater environment exploration, which can be observed in Section\ref{sec:ExperimentsResultsSynthetic} and \ref{sec:ExperimentsResultsReal-world}. Thus, we obtain a training set of 4000 ($1000\times 4$) samples, a validation set of 400 ($100\times 4$) samples, and a test set $\textbf{SyntTest}$ of 349 samples. For computational efficiency, we resize these images to $256\times 256$.

We trained the proposed JWCDN model using ADAM as the optimization algorithm with learning rate of 0.0002. The batch size is set to 1 and we trained the network for 100 epoches. All the parameters are chosen via cross-validation. We use pytorch as the deep learning framework on an Inter(R) i7-6700k CPU, 32GB RAM, and a Nvidia GTX 980 Ti GPU.

\subsection{Performance Evaluation Metrics}
To evaluate the capacity of the proposed method, we employ full-reference and no-reference image quality assessment techniques on synthetic underwater images and real-world underwater images respectively, since there are reference images for synthetic images, but not for real-world images.

\subsubsection{Full-reference Image Quality Assessments}
We use four full-reference image quality assessment techniques, i.e., MSE (Mean Squared Error), PSNR (Peak Signal Noise Ratio), SSIM (Structural Similarity Index) and Patch-based Contrast Quality Index (PCQI)\cite{Wang2015A}, to assess the performance of the proposed method. Generally, MSE and PSNR are used to evaluate image noise, where lower MSE and higher PSNR values indicate less noise. SSIM measures the visual impact of three characteristics of an image: luminance, contrast and structure. A higher SSIM value denotes a better result of enhancement. PCQI measures the contrast of images, and a higher PCQI indicates that the image has better contrast.

\subsubsection{No-reference Image Quality Assessments}

Additionally, we use two no-reference image quality assessment techniques, i.e., Blur Metric\cite{Crete2007The} and underwater image quality measure (UIQM)\cite{Panetta2016Human}, to assess the performance of the proposed method. Blur Metric is utilized in evaluating the blur effect of an image. It ranges from 0 to 1 which are respectively the best and the worst quality in term of blur perception. UIQM is a linear combination of three independent image quality measures
\begin{eqnarray}
\mathrm{UIQM} = \epsilon_{1}\times \mathrm{UICM} + \epsilon_{2}\times \mathrm{UISM} + \epsilon_{3} \times \mathrm{UIConM}
\label{Eqs-UIQM}
\end{eqnarray}
where UICM denotes the colorfulness, UISM denotes the sharpness, and UIConM denotes the contrast measures. The parameters $\epsilon_{1}$, $\epsilon_{2}$ and $\epsilon_{3}$ are weights, whose values are application dependent. In this paper, the values are as follows: $\epsilon_{1} = 0.3282$, $\epsilon_{2} = 0.2953$, and $\epsilon_{3} = 3.5753$. A greater UIQM value indicates superior image quality.

\begin{figure}[!t]
\centering
\includegraphics[width=3.5in]{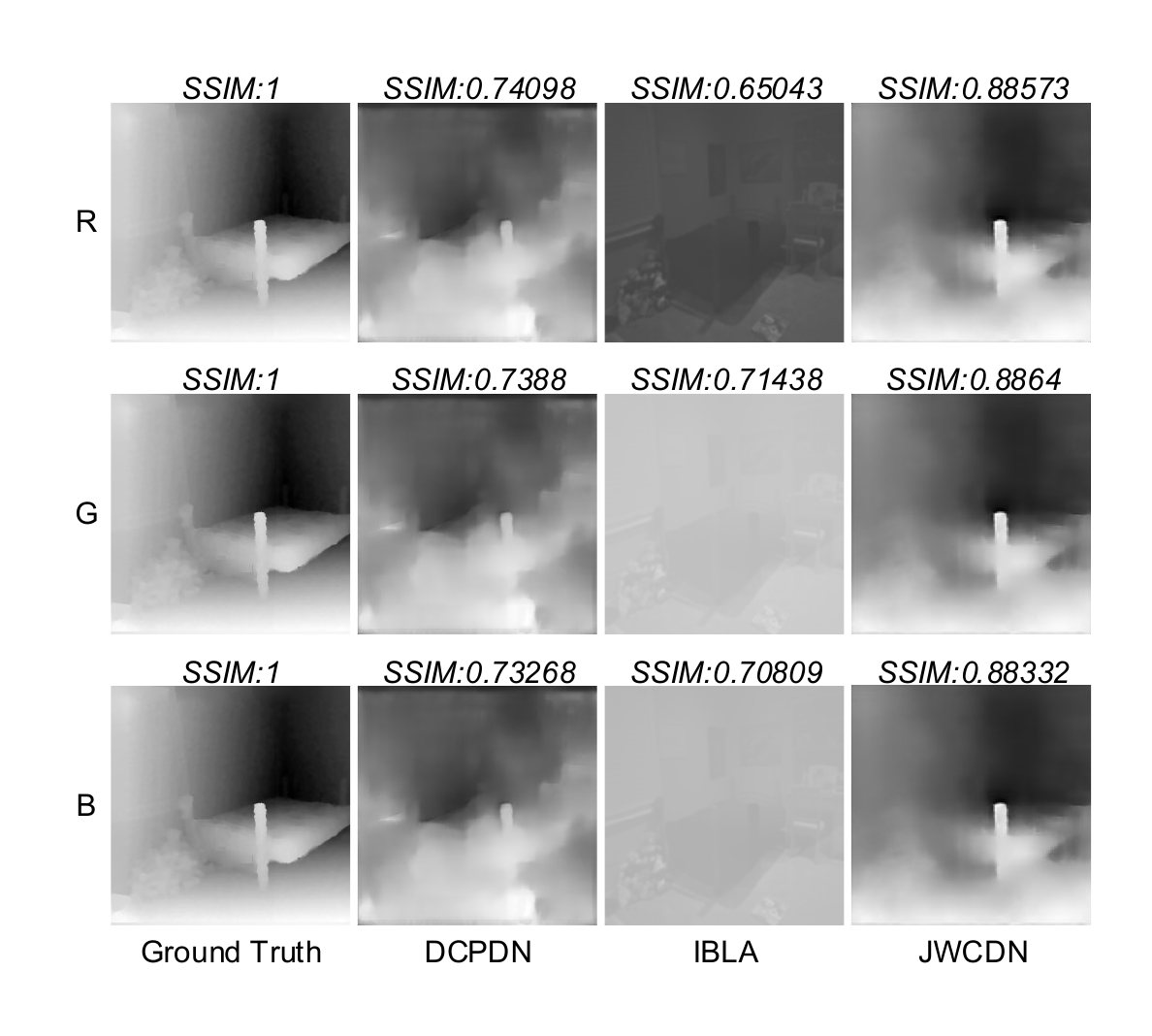}
\caption{Transmission map estimation results. $R$, $G$ and $B$ denotes red, green and blue channels respectively.}
\label{fig_trans}
\end{figure}

\begin{figure*}[!t]
\centering
\includegraphics[width=7.1in]{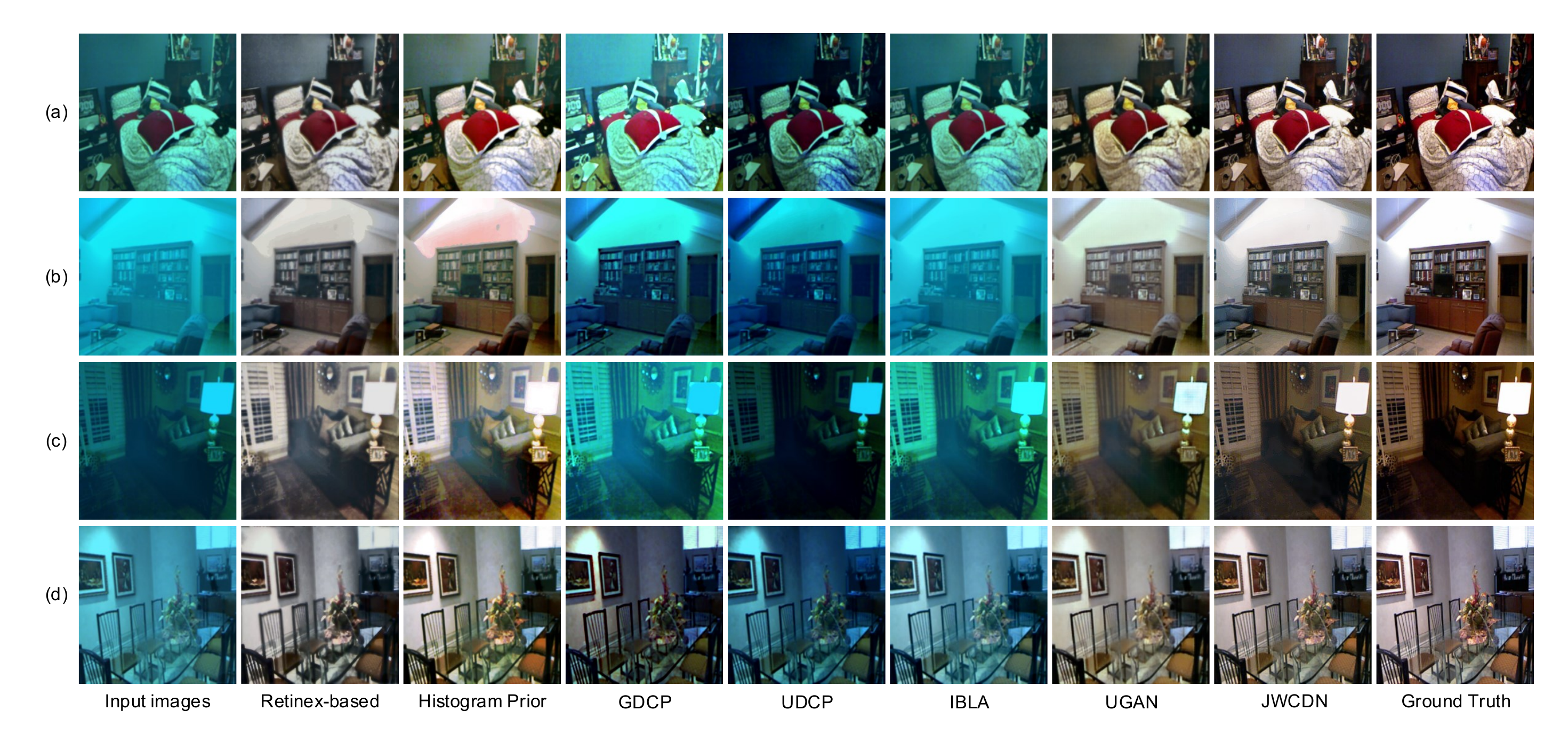}
\caption{Qualitative comparisons for samples from the synthetic test dataset $\textbf{SyntTest}$. (a)-(d) represent the samples selected from $\textbf{SyntTest}$.}
\label{fig_synt}
\end{figure*}

\subsection{Transmission Map Estimation Results}
Fig. \ref{fig_trans} shows the comparisons of transmission map results estimated from synthetic underwater images. we compare our results with DCPDN \cite{Zhang2018Densely} and IBLA, which compute the transmission map of RGB channel. We don't compare with Histogram Prior which also estimates the transmission map of RGB channel, due to the source code provided by the authors cannot output the transmission map. DCPDN is a state-of-the-art dehazing method that inspired us to embed physical model into networks. The poor transmission maps predict by DCPDN show that color casts caused by wavelength attenuation lead to a challenge for transmission map estimation. The transmission maps compute by IBLA lack contrast and edges. Fig. \ref{fig_trans} also shows the SSIM values of these methods. The SSIM values of our method are higher than other methods. It is noteworthy that transmission map and other parameters affect the final result simultaneously for those hazy image formation model based methods. Thus, a better transmission map do not always mean a better enhanced result. Hence, we concentrate on the final enhanced results of methods.

\renewcommand\arraystretch{1.2}
\begin{table}[]
\scriptsize
\centering
\caption{Quantitative results of samples in Fig. \ref{fig_synt} evaluated by full-reference image quality assessments.}
\label{Tab_synt}
\begin{tabular}{cccccc}
\toprule
Images& Methods& MSE & PSNR & SSIM & PCQI  \\
\midrule
\multirow{6}{*}{\shortstack{Fig.\ref{fig_synt} (a)}}
& Retinex-based	&	0.75428	&	19.3555	&	0.68943	&	0.60774	\\
& Histogram Prior	&	0.83377	&	18.9203	&	0.7056	&	0.69396	\\
& GDCP	&	6.461	&	10.0278	&	0.19104	&	0.63239	\\
& UDCP	&	2.9314	&	13.46	&	0.30212	&	0.57051	\\
& IBLA	&	2.1798	&	14.7467	&	0.24414	&	0.62717	\\
& UGAN	&	0.48124	&	21.3072	&	0.73394	&	0.64182	\\
& JWCDN	&	\textbf{0.1919}	&	\textbf{25.2999}	&	\textbf{0.90599}	&	\textbf{0.85296}	\\
\midrule
\multirow{6}{*}{\shortstack{Fig.\ref{fig_synt} (b)}}
& Retinex-based	&	0.92176	&	18.4846	&	0.65093	&	0.63831	\\
& Histogram Prior	&	0.71568	&	19.5836	&	0.61086	&	0.69222	\\
& GDCP	&	5.7757	&	10.5148	&	-0.004177	&	0.58308	\\
& UDCP	&	8.2103	&	8.9872	&	-0.059201	&	0.4776	\\
& IBLA	&	7.8719	&	9.17	&	-0.09394	&	0.49143	\\
& UGAN	&	0.88448	&	18.6639	&	0.60331	&	0.6498	\\
& JWCDN	&	\textbf{0.37774}	&	\textbf{22.3589}	&	\textbf{0.8342}	&	\textbf{0.73362}	\\
\midrule
\multirow{6}{*}{\shortstack{Fig.\ref{fig_synt} (c)}}
& Retinex-based	&	4.9681	&	11.1689	&	0.3251	&	0.70255	\\
& Histogram Prior	&	6.7705	&	9.8246	&	0.34668	&	0.74948	\\
& GDCP	&	8.3907	&	8.8928	&	-0.17418	&	0.68598	\\
& UDCP	&	2.173	&	14.7601	&	-0.066646	&	0.531	\\
& IBLA	&	5.51	&	10.7193	&	-0.2042	&	0.67729	\\
& UGAN	&	1.4321	&	16.5711	&	0.52122	&	0.6418	\\
& JWCDN	&	\textbf{0.53028}	&	\textbf{20.8857}	&	\textbf{0.66189}	&	\textbf{0.72393}	\\
\midrule
\multirow{6}{*}{\shortstack{Fig.\ref{fig_synt} (d)}}
& Retinex-based	&	0.67011	&	19.8693	&	0.71964	&	0.64081	\\
& Histogram Prior	&	0.32374	&	23.0288	&	0.76858	&	0.7755	\\
& GDCP	&	3.0166	&	13.3356	&	0.52476	&	0.61137	\\
& UDCP	&	4.073	&	12.0317	&	0.20753	&	0.50985	\\
& IBLA	&	1.4268	&	16.5872	&	0.23873	&	0.66525	\\
& UGAN	&	0.28105	&	23.643	&	0.77105	&	0.68659	\\
& JWCDN	&	\textbf{0.1156}	&	\textbf{27.5011}	&	\textbf{0.924}	&	\textbf{0.83428}	\\
\bottomrule
\end{tabular}
\end{table}

\renewcommand\arraystretch{1.2}
\begin{table}[]
\scriptsize
\centering
\caption{Quantitative results evaluated on synthetic $\textbf{SyntTest}$ dataset by full-reference image quality assessments.}
\label{Tab_synt_set}
\begin{tabular}{ccccc}
\toprule
Methods& MSE & PSNR & SSIM & PCQI  \\
\midrule
Retinex-based	&	1.7625	&	16.7678	&	0.53271	&	0.6552	\\
Histogram Prior	&	1.9222	&	16.9358	&	0.60874	&	0.73271	\\
GDCP	       &	7.4526	&	10.01	&	-0.0077371	&	0.60367	\\
UDCP	       &	5.4809	&	11.3138	&	0.017639	&	0.5326	\\
IBLA	      &	5.2885	&	11.4254	&	-0.065604	&	0.58303	\\
UGAN	       &	1.1734	&	18.5	&	0.59515	&	0.64133	\\
JWCDN	      &	\textbf{1.0125}	&	\textbf{21.096}	&	\textbf{0.75712}	&	\textbf{0.76362}	\\
\bottomrule
\end{tabular}
\end{table}
\subsection{Experiments Results for Synthetic Underwater Images}\label{sec:ExperimentsResultsSynthetic}
We evaluate the capacity of the JWCDN model on the synthetic dataset $\textbf{SyntTest}$, comparing with several state-of-the-art methods. Fig. \ref{fig_synt} shows the results of underwater image enhancement on four sample images from the synthetic dataset $\textbf{SyntTest}$.

As shown in Fig. \ref{fig_synt}, GDCP\cite{Peng2018Generalization}, UDCP\cite{Drews2016Underwater} and IBLA\cite{Peng2017Underwater} perform under enhancement in color compensation, while Retinex-based\cite{Fu2014Retinex} and Histogram Prior\cite{Li2016Underwater} perform over enhancement in color compensation. UGAN\cite{Fabbri2018Enhancing} and the proposed JWCDN provide a better color appearance. Comparing UGAN with JWCDN, it can be observed that JWCDN preserve sharper contours with less color distortion and are more visually closer to the ground-truth. Hence, even though these methods are able to enhance the input images, they tend to either over enhance or under enhance the images.

Since the dataset $\textbf{SyntTest}$ is synthesized, the ground truth images are available, enabling us to evaluate the performance quantitatively with full-reference image quality assessments. We evaluate all samples in $\textbf{SyntTest}$ by four full-reference image quality assessment techniques, i.e. MSE, PSNR, SSIM and PCQI. The quantitative results of samples shown in Fig. \ref{fig_synt} are tabulated in Table \ref{Tab_synt}. And the average quantitative results evaluated on $\textbf{SyntTest}$ are tabulated in Table \ref{Tab_synt_set}. The values in bold represent the best results. In Table \ref{Tab_synt} and Table \ref{Tab_synt_set}, our method obtains the lower MSE, higher PSNR, SSIM and PCQI values than other methods. The lower MSE and higher PSNR values mean that our results have less noise. And the higher SSIM and PCQI values show that our method can maintain structure and improve contrast.

\subsection{Experiments Results for Real-world Underwater Images}\label{sec:ExperimentsResultsReal-world}

To demonstrate the generalization ability of the proposed method, we evaluate the proposed method on two real-world underwater image datasets $\textbf{RealA}$ and $\textbf{RealB}$. We gather underwater images provided by previous methods and downloaded from the Internet to establish dataset $\textbf{RealA}$. In addition, we establish dataset $\textbf{RealB}$ with seven underwater videos provided by National Natural Science Foundation of China (NSFC) and captured by ourselves. $\textbf{RealA}$ and $\textbf{RealB}$ contain 178 and 6778 images respectively. These real-world underwater images are characterized by different tones, lights, and contrasts. The quantitative evaluation on real-world underwater images is performed by no-reference image quality assessment techniques, i.e. Blur Metric and UIQM, due to the lack of reference images.

\begin{figure*}[!t]
\centering
\includegraphics[width=7.1in]{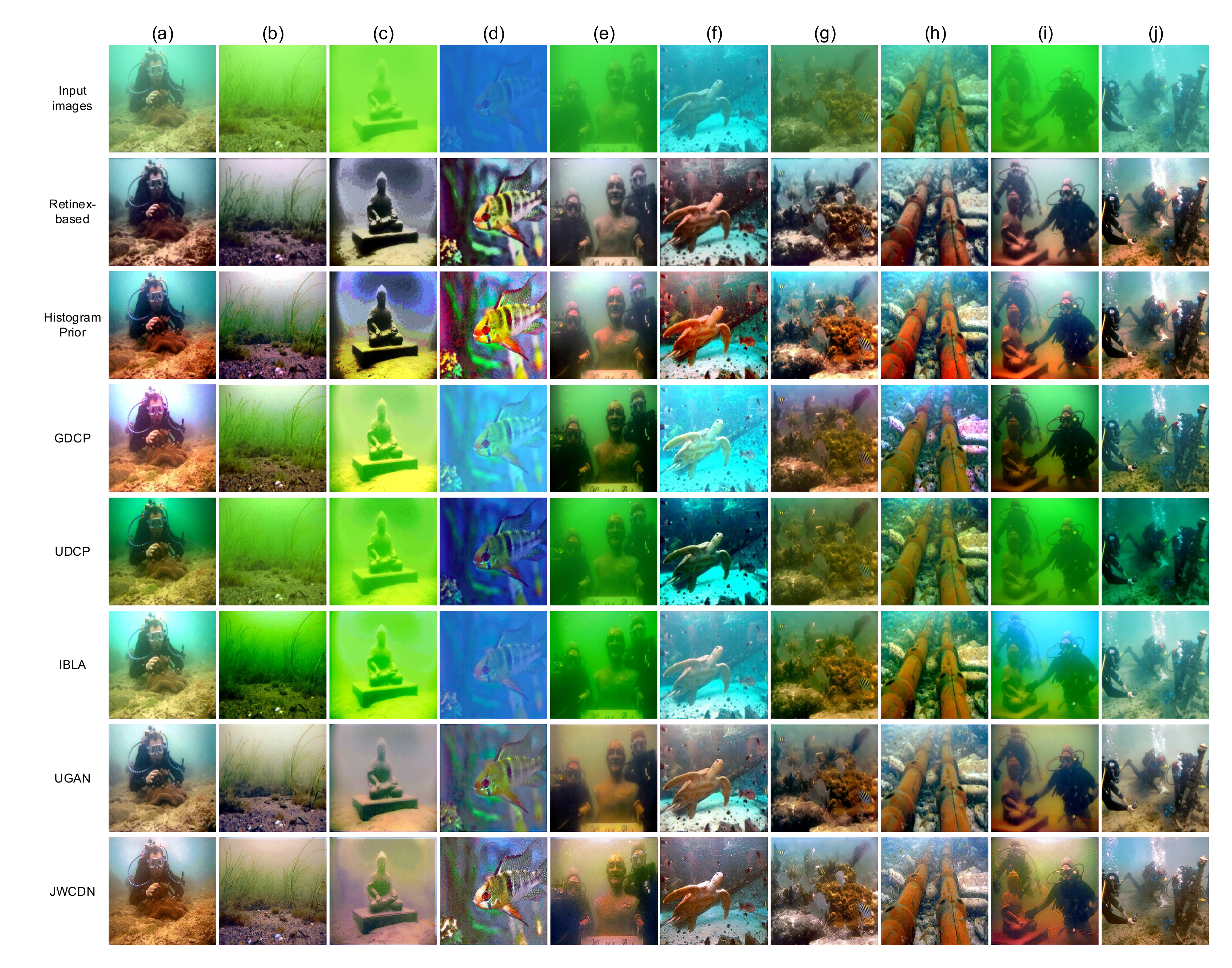}
\caption{Qualitative comparisons for samples from the real-world image dataset $\textbf{RealA}$. (a)-(j) represent the samples selected from $\textbf{RealA}$.}
\label{fig_real}
\end{figure*}

Fig. \ref{fig_real} presents the visual comparisons with competitive methods on ten sample images from the dataset $\textbf{RealA}$. A first glance at Fig. \ref{fig_real} may give the impression that the results of Retinex-based and Histogram Prior might be more bright-colored; however, a careful inspection reveals that the Retinex-based method leads to a darker appearance. Similarly, Histogram Prior method causes over-saturation that some results have red tone. And the images produced by the GDCP consist of over-enhancement and under-saturation. The methods of UDCP and IBLA have little effect on the inputs. UGAN and the proposed JWCDN model show an approximate tone on different real-world images. But JWCDN causes more sharper details. We evaluate all samples in $\textbf{RealA}$ by Blur Metric and UIQM. The quantitative results of samples shown in Fig. \ref{fig_real} are tabulated in Table \ref{Tab_real}. It can be observed that our method obtains almost lower Blur Metric values and higher UIQM values compared with competitive methods. Hence, our method can produce better colorfulness, sharpness and contrast performance on real-world underwater images from a objective perspective.

\renewcommand\arraystretch{1.2}
\begin{table*}[]
\scriptsize
\centering
\caption{Quantitative blur metric and UIQM values of samples in Fig. \ref{fig_real}.}
\label{Tab_real}
\begin{tabular}{ccccccccccccc}
\toprule
Assessments& Methods & Fig.\ref{fig_real}(a) &Fig.\ref{fig_real}(b) &Fig.\ref{fig_real}(c) &Fig.\ref{fig_real}(d) &Fig.\ref{fig_real}(e) &Fig.\ref{fig_real}(f) & Fig.\ref{fig_real}(g) &Fig.\ref{fig_real}(h) &Fig.\ref{fig_real}(i) &Fig.\ref{fig_real}(j) \\
\midrule
\multirow{7}{*}{\shortstack{Blur Metric}}
& Retinex-based   & 0.37846&	0.35735&	0.38729& 0.42717&	0.48342&	0.40059&	0.39448&	0.43059&	0.50443&	0.39479\\
& Histogram Prior & 0.28499&	0.25783&	\textbf{0.27234}&	0.24777&	0.25985&	0.30177&	0.31817&	0.36433&	0.30013&	0.29224\\
& GDCP            & 0.28401&	0.25649	&	0.38907	&	0.29544	&	0.28975	&	0.27399	&	0.3227	&	0.35041	&	0.31127	&	0.27414\\
& UDCP            & 0.31313	&	0.29685	&	0.43346	&	0.37774	&	0.37516	&	0.32899	&	0.34827	&	0.3671	&	0.40191	&	0.32393\\
& IBLA           & 0.32288	&	0.28918	&	0.48859	&	0.36555	&	0.38592	&	0.33763	&	0.3575	&	0.37457	&	0.44518	&	0.34155\\
& UGAN            & 0.34443	&	0.33148	&	0.39659	&	0.36609	&	0.37244	&	0.36514	&	0.39748	&	0.40695	&	0.38122	&	0.38265\\
& JWCDN           & \textbf{0.21917}&	\textbf{0.21552}	&0.30723	&	\textbf{0.21007}	&	\textbf{0.22409}	&	\textbf{0.21637}	&	\textbf{0.24649}	&	\textbf{0.25254}	&	\textbf{0.23204}	&	\textbf{0.22668}\\
\midrule
\multirow{7}{*}{\shortstack{UIQM}}
& Retinex-based	&	1.4082	&	1.3606	&	1.4139	&	1.5238	&	1.0301	&	1.4679	&	1.4992	&	1.5011	&	1.1794	&	1.3998 \\
& Histogram Prior	&	1.4873	&	1.5782	&	\textbf{1.6481}	&	1.7725	&	1.1452	&	\textbf{1.7054}	&	1.566	&	1.5731	&	1.2646	&	1.5021 \\
& GDCP	&	1.3717	&	1.5007	&	0.92029	&	0.8883	&	1.343	&	1.2115	&	1.3788	&	1.5374	&	1.3314	&	1.3417 \\
& UDCP	&	1.3136	&	1.0781	&	0.74235	&	1.5095	&	0.76163	&	1.5629	&	1.2525	&	1.4199	&	0.79295	&	1.4576 \\
& IBLA	&	1.1702	&	1.496	&	0.77893	&	0.81798	&	0.85676	&	1.1652	&	1.3751	&	1.4954	&	0.90722	&	1.0924 \\
& UGAN	&	1.2492	&	1.3809	&	0.96145	&	1.1844	&	1.062	&	1.3013	&	1.3927	&	1.3832	&	1.1441	&	1.2587 \\
& JWCDN	&	\textbf{1.5368}	&	\textbf{1.5995}	&	1.1011	&	\textbf{1.8028}	&	\textbf{1.5415}	&	1.6791	&	\textbf{1.5754}	&	\textbf{1.6309}	&	\textbf{1.5139}	&	\textbf{1.5182} \\
\bottomrule
\end{tabular}
\end{table*}

\begin{figure*}[!t]
\centering
\includegraphics[width=7.1in]{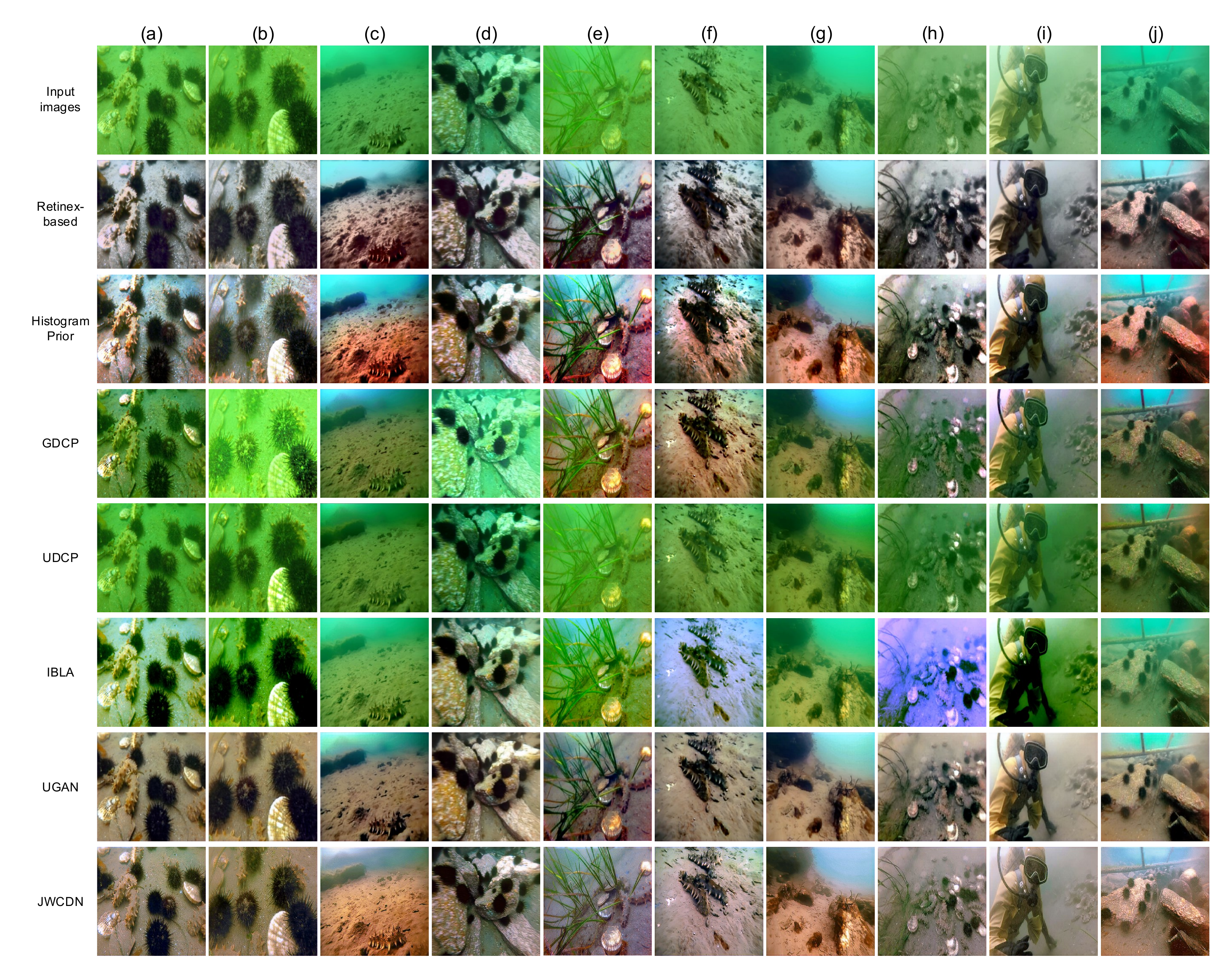}
\caption{Qualitative comparisons for samples from real-world image dataset $\textbf{RealB}$. (a)-(j) represent the samples selected from $\textbf{RealB}$.}
\label{fig_real_video}
\end{figure*}

Fig. \ref{fig_real_video} presents the results of underwater image enhancement on ten sample images from the dataset $\textbf{RealB}$. Similar to the enhanced results of $\textbf{RealA}$, the results of Retinex-based and Histogram Prior give the bright-coloured impression at first glance. However, the appearance of results provided by Retinex-based and Histogram Prior methods looks over-enhanced and under-saturated. GDCP, UDCP and IBLA have less effect on the inputs than other methods. Comparing with UGAN, JWCDN shows a similar color appearance but a better detail appearance. We also evaluate all samples in $\textbf{RealB}$ by Blur Metric and UIQM. Table \ref{Tab_real_video} shows the quantitative results of samples shown in Fig. \ref{fig_real_video}. It can be observed that our method obtains lowest Blur Metric values and higher UIQM values.

Furthermore, the average quantitative results evaluated on $\textbf{RealA}$ and $\textbf{RealB}$ are tabulated in Table \ref{Tab_real_set}. It can be observed that our method obtains best quantitative results on dataset $\textbf{RealA}$. And on dataset $\textbf{RealB}$, our method performs a best Blur Metric result and a second-highest UIQM result after Histogram Prior method. As visible, among all competitive methods we tested, JWCDN performs a better appearance across all assessments, demonstrating its effectiveness and robustness. And we do agree that the proposed method does not enhance real-world images as accurately as it does the synthetic ones, which can be improved by adding more underwater images in model training.

\renewcommand\arraystretch{1.2}
\begin{table*}[]
\scriptsize
\centering
\caption{Quantitative blur metric and UIQM values of samples in Fig. \ref{fig_real_video}.}
\label{Tab_real_video}
\begin{tabular}{ccccccccccccc}
\toprule
Assessments& Methods & Fig.\ref{fig_real_video}(a) &Fig.\ref{fig_real_video}(b) &Fig.\ref{fig_real_video}(c) &Fig.\ref{fig_real_video}(d) &Fig.\ref{fig_real_video}(e) &Fig.\ref{fig_real_video}(f) & Fig.\ref{fig_real_video}(g) &Fig.\ref{fig_real_video}(h) &Fig.\ref{fig_real_video}(i) &Fig.\ref{fig_real_video}(j) \\
\midrule
\multirow{7}{*}{\shortstack{Blur Metric}}
& Retinex-based	&	0.34445	&	0.34615	&	0.34514	&	0.49689	&	0.34674	&	0.36959	&	0.41908	&	0.45054	&	0.4746	&	0.37342\\
& Histogram Prior	&	0.27264	&	0.26037	&	0.27603	&	0.4452	&	0.2726	&	0.29985	&	0.32912	&	0.36924	&	0.38515	&	0.31061\\
& GDCP	&	0.26652	&	0.24205	&	0.2823	&	0.39672	&	0.27881	&	0.29999	&	0.32627	&	0.35688	&	0.36626	&	0.30562\\
& UDCP	&	0.29572	&	0.28789	&	0.29569	&	0.43593	&	0.28826	&	0.32146	&	0.33008	&	0.37808	&	0.39286	&	0.31773\\
&IBLA	&	0.2914	&	0.27212	&	0.29421	&	0.4575	&	0.26533	&	0.30164	&	0.34701	&	0.34833	&	0.3887	&	0.32331\\
& UGAN	&	0.39234	&	0.356	&	0.34734	&	0.48702	&	0.36779	&	0.35634	&	0.40308	&	0.42339	&	0.43162	&	0.42248\\
& JWCDN	&	\textbf{0.23142}	&	\textbf{0.21131}	&	\textbf{0.22145}	&	\textbf{0.32451}	&	\textbf{0.24037}	&	\textbf{0.23958}	&	\textbf{0.24082}	&	\textbf{0.27185}	&	\textbf{0.29438}	&	\textbf{0.21473}\\
\midrule
\multirow{7}{*}{\shortstack{UIQM}}
&Retinex-based	&	1.5302	&	1.4565	&	1.5054	&	1.4783	&	1.62	&	1.5305	&	1.2939	&	1.4186	&	1.2151	&	1.4107\\
& Histogram Prior	&	1.5679	&	1.4663	&	1.5763	&	1.512	&	\textbf{1.6985}	&	\textbf{1.6096}	&	1.4099	&	\textbf{1.505}	&	1.2233	&	1.4594\\
& GDCP	&	1.5597	&	1.3786	&	1.2945	&	1.3905	&	1.5905	&	1.5937	&	1.2011	&	1.3603	&	1.1536	&	1.3938\\
& UDCP	&	1.3067	&	1.3905	&	1.1487	&	1.4779	&	1.1034	&	1.2145	&	1.1263	&	1.1317	&	1.0004	&	1.3138\\
& IBLA	&	1.5647	&	\textbf{1.6369}	&	1.1169	&	1.46	&	1.5665	&	1.3531	&	1.0911	&	1.2896	&	\textbf{1.3867}	&	1.1118\\
& UGAN	&	1.3909	&	1.3715	&	1.3498	&	1.4035	&	1.4326	&	1.3111	&	1.2819	&	1.2927	&	1.091	&	1.2517\\
& JWCDN	&	\textbf{1.5845}	&	1.577	&	\textbf{1.58}	&	\textbf{1.6269}	&	1.5893	&	1.5951	&	\textbf{1.5269}	&	1.4273	&	1.1759	&	\textbf{1.5372}\\
\bottomrule
\end{tabular}
\end{table*}

\renewcommand\arraystretch{1.2}
\begin{table*}[]
\scriptsize
\centering
\caption{Average quantitative blur metric and UIQM values on real-world datasets $\textbf{RealA}$ and $\textbf{RealB}$.}
\label{Tab_real_set}
\begin{tabular}{ccccccccc}
\toprule
Datasets & Assessments & Retinex-based & Histogram Prior & GDCP & UDCP & IBLA & UGAN & JWCDN \\
\midrule
\multirow{2}{*}{\shortstack{$\textbf{RealA}$}}
&Blur Metric & 0.37958 & 0.26556 & 0.2908 & 0.33736 & 0.34485 & 0.36843 & \textbf{0.21818}\\
\cmidrule(r){2-9}
&UIQM & 1.3962 & 1.5315 & 1.3979 & 1.4161 & 1.2352 & 1.2894 & \textbf{1.574}\\
\midrule
\multirow{2}{*}{\shortstack{$\textbf{RealB}$}}
&Blur Metric & 0.3944 & 0.31418 & 0.31795 & 0.3323 & 0.33181 & 0.40356 & \textbf{0.25059}\\
\cmidrule(r){2-9}
&UIQM & 1.3838 & \textbf{1.4688} & 1.2882 & 1.1888 & 1.2192 & 1.2359 & 1.4605\\
\bottomrule
\end{tabular}
\end{table*}

\subsection{Ablation Study}

To demonstrate the effect of each module in proposed network, we carry out an ablation study involving the following experiments:

1) JWCDN without content preserving loss (JWCDN-woCL),

2) JWCDN without edge preserving module (JWCDN-woEP),

3) JWCDN.

\begin{figure}[!t]
\centering
\includegraphics[width=3.4in]{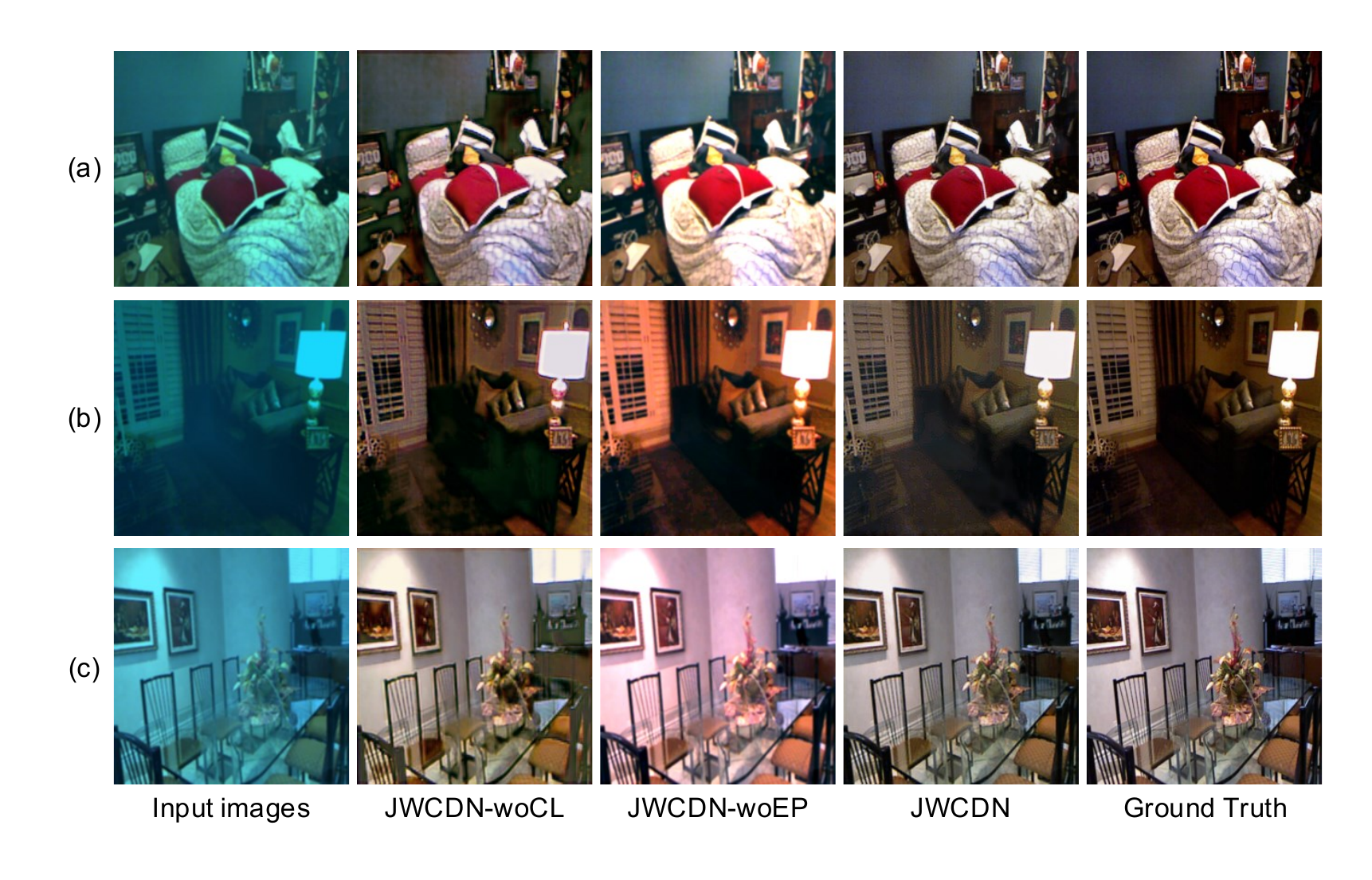}
\caption{Enhanced results using different modules.}
\label{fig_ablation}
\end{figure}
The evaluation is performed on the synthesized $\textbf{TestA}$ datasets. Visual comparisons are shown in the Fig. \ref{fig_ablation}. From Fig. \ref{fig_ablation}, we make the following observations: 1) The results of JWCDN-woCL show a blurriness appearance compared with JWCDN. The use of content preserving loss is able to better preserve edges. 2) JWCDN-woEP leads to a red tone, and produces some over bright regions. And comparing with JWCDN, the results of JWCDN-woEP are also perform some blurriness. Hence, the fine-scale module is able to refine the brightness and preserve the edges.

\begin{table}[]
\scriptsize
\centering
\caption{Average quantitative results for ablation study evaluated on synthetic $\textbf{SyntTest}$ dataset.}
\label{Tab_ablation}
\begin{tabular}{ccccc}
\toprule
Methods& MSE & PSNR & SSIM & PCQI  \\
\midrule
JWCDN-woCL & 1.1898 & 18.8177 & 0.63553 & 0.64072\\						
JWCDN-woFS & 8.9867 & 10.9361 & 0.44745 & 0.58658\\				
JWCDN & \textbf{1.0125} & \textbf{21.096} & \textbf{0.75712} & \textbf{0.76362}\\
\bottomrule
\end{tabular}
\end{table}

The full-reference assessment results averaged on enhanced images for the various configurations are tabulated in Table \ref{Tab_ablation}. The quantitative performance evaluated on $\textbf{TestA}$ also demonstrates the effectiveness of each module.

\section{Conclusion}
We propose a novel jointly wavelength compensation and dehazing network that can jointly learn the transmission map, wavelength attenuation and background light via different network modules, and recover underwater image all together. The proposed method is achieved by directly embedding underwater formation model into the generative adversarial network, considering wavelength attenuation along surface-object path and the scattering along object-camera path in the water medium simultaneously. To efficiently estimate transmission map, a multi-scale densely connected encoder-decoder network with a gradient preserving loss function is proposed to leverage features from multiple layers. Additionally, we propose a novel underwater image synthesis method based on inherent optical properties of different water types, to simulate the color and blurriness appearance of real-world underwater environments. Extensive experiments are conducted to show the significance of the proposed method. In future, we plan to improve our current model to limit the over-saturation which occurs in some experimental results. And we will develop our current underwater image synthesis method to generate images which are closer to real-world underwater images.




\ifCLASSOPTIONcaptionsoff
  \newpage
\fi

%








\end{document}